\documentclass[conference]{IEEEtran}
\IEEEoverridecommandlockouts

\usepackage{cite}
\usepackage{amsmath,amssymb,amsfonts}
\usepackage{algorithmic}
\usepackage{graphicx}
\usepackage{url}
\usepackage{textcomp}
\usepackage{xcolor}
\def\BibTeX{{\rm B\kern-.05em{\sc i\kern-.025em b}\kern-.08em
    T\kern-.1667em\lower.7ex\hbox{E}\kern-.125emX}}
    
\begin{document}

\title{Democracy for DAOs: An Empirical Study of Decentralized Governance and Dynamics\\
{\Large Case Study Internet Computer SNS Ecosystem}}

\author{
\IEEEauthorblockN{Burak Arda Okutan}
\IEEEauthorblockA{\textit{TU Berlin, Germany} }
\and
\IEEEauthorblockN{Stefan Schmid}
\IEEEauthorblockA{\textit{TU Berlin and Weizenbaum Institute, Germany
} }
\and
\IEEEauthorblockN{Yvonne-Anne Pignolet}
\IEEEauthorblockA{\textit{DFINITY, Switzerland} }
}


\maketitle
\let\thefootnote\relax\footnotetext{~Research in part supported by German Research Foundation (DFG), SPP 2378 (project ReNO), 2023--2027.}

\begin{abstract}
Decentralized autonomous organizations (DAOs) rely on governance mechanism without centralized leadership.
This paper presents an empirical study of user behavior in governance for a variety of DAOs, ranging from DeFi to gaming, using the  Internet Computer Protocol DAO framework called SNS (Service Nervous System). 

To analyse user engagement, we measure participation rates and
frequency of proposals submission and 
voter approval rates. 
We evaluate decision duration times to determine DAO agility.
To investigate dynamic aspects, we also measure metric shifts in time.
We evaluate over 3,000 proposals submitted in a time frame of 20 months from 14 SNS DAOs. The selected DAO have been existing between 6 and 20 months and  
cover a wide spectrum of use cases, treasury sizes, and
number of participants. 
We also compare our results for SNS DAOs with DAOs from other blockchain platforms. While approval rates are generally high for all DAOs studied, SNS DAOs show slightly more alignment. We observe that the SNS governance mechanisms and processes in ICP lead to higher activity, lower costs and faster decisions. 
Most importantly, in contrast to studies which report a decline in participation over time for other frameworks, SNS DAOs exhibit sustained or increasing engagement levels over time. 
\end{abstract}

\section{Introduction}
In traditional centralized systems, decision-making is often non-transparent and driven by narrow interests of a central authority, which may undermine fairness and trust in the system. 
With the advent of Web3, transparency, scalability, and community-driven decision-making gained interest and liquid democracy emerge as innovative mechanisms that address these challenges. Distributing decision power across broader communities ensures that decisions are more representative of the entire community's interests, preventing the influence of a central authority or powerful stakeholders. By supporting democratic voting schemes, DAOs can empower their community to 
shape the future development of the DAO, with voting on proposals encompassing a wide range of topics, including code changes, tokenomics, and other matters. Decentralized governance systems should aim at efficient decision-making, without overburdening voters, and therefore handle issues like voter fatigue. 

In recent years, liquid democracy and DAOs have gained much interest because of their potential to alter the existing governance models. Liquid Democracy combines elements of both direct and representative democracy \cite{swierczek2011}, giving participants a choice of voting directly or delegating decisions. The main benefits are the ability to balance voter fatigue and representation issues, thus enhancing the efficiency of decision-making processes. 

A survey on different governance models used in decentralized networks can be found in \cite{karjalainen2020}. Theoretical foundations, applications and adaptations of DAOs have been studied in \cite{qin2023,sims2019,wang2019}. 
 The level of decentralization has been investigated across various blockchain ecosystems and DAOs, including \cite{barbereau2022,kling2015,sun2022,fritsch2022,liu2024,schmid2024}.
Far less attention was devoted to other metrics, such as activity, approval rates or durations, even though theoretical studies pointed out that these aspects are critical to keep participants involved in their organization's activities and development \cite{wright2021}.

In this paper, we take an empirical approach and investigate how
efficient and active existing voting systems are, 
evaluating participation rates and engagement levels of users,
as well as 
activity levels within each community and frequency of voting proposals.
The governance participation rate measures the ratio of the
voting power of the users participating in a proposal to
the total voting power. 
Calculating average monthly proposal submission frequencies sheds light
on the activity level of each DAO.
This activity is likely influenced by the communities’ interests, its need for
development and management as well as governance structure and functionality.
Other factors that can influence the activity level are the resources participants must devote to governance, with respect to the time, expertise and monetary expenses, both for submitting and voting on proposals. 

We also study voter approval rates and decision duration times: 
a high approval rate and fast decision times 
may indicate a high trust and alignment in the community,
and/or a high degree of centralization.
Examining decision-making
durations further provides insight into the agility and inclusiveness
of each DAO’s governance processes, as well as its capacity
to handle a high volume of proposals, which is crucial, as the
scale and adoption of DAOs grow rapidly.
In order to shed light on trends, we 
are particularly interested in the evolution of the governance 
over time. 

For our study, we consider the Internet Computer Protocol (ICP) DAO framework called 
Service Nervous Systems (SNS)~\cite{assmann2023fully}, 
which features an innovative governance model with a particularly
flexible incentivized liquid democracy model, where votes can be delegated in complex ways. SNS DAOs extend the governance model of ICP’s Network Nervous System (NNS) to
Decentralized Application (dApps) governed by smaller communities with a specific
focus and aim. 

We collected data from over 3,000 different proposals by 14 SNSs, covering different ages, treasury size, and number of participants, as well as a variety of DAO use cases, which range from 
decentralized finance (DeFi) to gaming. 

\textbf{Our Contributions}
We present a large-scale empirical study to
shed light on the efficiency, scalability, and agility
of different SNS DAOs' governance processes on ICP. 

Average participation rate for the SNS DAOs is approximately 64\%, with Nuance, a publishing DAO, exhibiting the highest participation rate at
80\%, indicating a robust community engagement.
Across all DAOs, we observe
high approval rates, with most DAOs showing rates
above 90\%.
The
average decision-making duration for SNS DAOs is approximately 1.14 days (approximately 27 hours). We further find that decision-making durations
varies with proposal types, since different rules and mechanisms
are implemented for each of them and the complexity
of the decision-making as well as interest of  are also
influenced by it. 
In terms of dynamics, we observe small fluctuations and find that 
 participation rates show
an overall increase; approval rates
are more consistent.

We also compare our results for SNS DAOs with DAOs from other blockchain platforms, in particular Ethereum-based DAOs like Compound, UniSwap, ENS, and Gitcoin. We find that governance mechanisms and processes in ICP are mostly more efficient in terms of proposal activity, community engagement and decision-making durations and costs. While approval rates are generally high for the DAOs studied, SNS DAOs have a slightly higher consensus.

\textbf{Paper Organization}
The remainder of this paper is organized as follows. We introduce the necessary background in Section~\ref{sec:background}. 
We report on our methodology and empirical results in Section~\ref{sec:eval} and compare them to DAOs on other blockchain platforms in Section~\ref{sec:comparison}, concluding in Section~\ref{sec:conclusion}.


\section{Background}\label{sec:background}

\subsection{Internet Computer Protocol (ICP)}
Internet Computer Protocol (ICP) aims to provide efficient multi-tenant, general-purpose, and secure compute - capable to host software ranging from social media, enterprise applications, or personal websites to AI applications such as chat bots. In contrast to many other compute platforms, ICP is decentralized which means that it runs on many node machines distributed around the world and maintained by independent node providers. ICP coordinates these node machines to execute smart contracts called canisters building on blockchain technology, often at a fraction of the cost of other blockchains. The ICP whitepaper~\cite{dfinity2022} provides a more comprehensive introduction.

The Internet Computer utilizes a \emph{reverse gas model}.
Instead of users bearing the computational costs, developers
pre-charge dapps by burning ICP tokens to pay for the
dapp’s computations and storage. Thus users engaging with dapps are
not burdened with transaction fees, and they do not require
specialized wallets.

\textbf{Network Nervous System (NNS).} A special set of canisters, the Network Nervous System (NNS), manages ICP's governance~\cite{assmann2023fully}. ICP token holders can lock some token in \emph{Neurons} that then grant them voting rights to influence the evolution of the platform with respect to the protocol versions, network topology,  node provider remuneration etc. To this end, the ICP token holders interact with the governance canister to submit proposals and vote on them. Once a proposal has been accepted, it is executed autonomously based on pre-defined rules in the canisters.  For example, a proposal could change the amount node providers receive for their services or deploy a new version of the underlying protocol.
The voting power of a neuron depends on the token amount, their locking period and age, with rewards proportional to voting power to incentivize long-term engagement.
Neuron owners may not have the time or necessary expertise to evaluate each individual proposal, which could hinder voter engagement. However, Neuron owners can delegate their voting power to other neurons they trust, implementing a hybrid form of liquid democracy. Instead of assigning the voting power to a single other neuron,  automatic voting can be configured by \emph{following} the voting decision made by a group of other neurons, called followees. Proposals are grouped into different proposal topics and following is done on a per-topic basis. A neuron casts a vote if a majority of its followees agree; otherwise, it abstains.\footnote{\url{https://internetcomputer.org/docs/current/developer-docs/daos/nns/concepts/neurons/neuron-following}} 
The integration of this form of liquid democracy provides unique opportunities regarding voter engagement and fatigue \cite{schmid2024}, \cite{liu2024}. The reverse gas model benefits NNS DAO members, enabling them to vote on numerous proposals at no cost.

\textbf{Service Nervous System (SNS).}
SNS DAOs extend the NNS governance model to individual decentralized applications, enabling developers to adopt a shared, flexible DAO structure. Instead of implementing their own governance structures for every project, the SNS DAO framework can be used, so the application developers can focus more on the development of the projects. The SNS governance system is designed to be highly flexible, allowing each dapp to design its governance model according to community needs via parametrization e.g., for the voting reward percentage and by adding additional application-specific proposals. In other words, all SNSs use the same canister code, while communities configure governance rules, tokenomics, and proposal types based on their needs.

SNS DAOs offer fully on-chain governance, minimal costs, and advanced delegation. The reverse gas model shifts transaction fees from users to developers, lowering proposal costs to ~\$11 compared to thousands on Ethereum, with high activity levels observed in this paper supporting this factor. All governance interactions occur on-chain unlike many Ethereum DAOs, enhancing transparency. Each SNS includes a ledger canister that defines a unique token for each SNS, which can be locked in a Neuron to participate in voting on the evolution of the SNS via the governance canister. Neurons are rewarded proportionally to activity and commitment, incentivizing sustained participation. SNSs hold treasuries of ICP and SNS tokens for operations like development and marketing. As in the NNS, neurons can vote directly or via delegation. Topic-specific majority delegation improves inclusivity and efficiency. These features make SNS DAOs an effective model for studying scalable, incentive-aligned decentralized governance.

The proposal data offered by the governance canister of an SNS contains attributes such as submission and decision timestamps, voting decisions ("Yes"/"No" votes in voting power and "total" voting power) as well as information on proposal types and action IDs, which can be used for classifying the proposals in different categories.

A proposal can be decided in two ways; either by an absolute majority -more than half of the total voting power stored in the proposal- before the voting period ends or a simple majority -more than half of the cast votes with at least achieves 3\% of the total voting power- at the end of the voting period. Critical proposals which transfer tokens from the treasury require longer voting periods and higher thresholds. For instance, the voting period for critical proposal types is 5-10 days while for non-critical proposals the default is 4-8 days.

\section{Methodology and Evaluation}\label{sec:eval}

\subsection{Data Collection and Evaluation}
For evaluating governance dynamics within SNS DAOs, data was collected for 14 out of a total of 29 SNS DAOs. We selected a subset of DAOs with a minimum age of 6 months to cover a wide range of categories, age, treasury size and number of participants, see Table~\ref{table:DAO_overview}. 

\begin{table}[t]\vspace{-.4cm}
\caption{List of SNS DAOs}
\begin{center}
\begin{tabular}{c c c c r}
\textbf{ \hspace{0.2cm} Name \hspace{0.2cm}} & \textbf{Age} & \textbf{Category} & \textbf{\hspace{0.1cm} Treasury\hspace{0.1cm}} & \textbf{\hspace{0.2cm}Neurons} \\
& months & &\hspace{0.1cm} USD \hspace{0.1cm} & \\
\hline
\end{tabular}
\begin{tabular}{c c c r r}
OpenChat & 20 & Chat &  \$ 26,453,292 & 29,497 \\
ICLighthouse & 7 & DeFi &  \$ 41,721,648 & 7,108 \\
ICPSwap & 6 & DeFi &  \$ 11,748,411 & 10,220 \\
SONIC & 12 & DeFi &  \$ 7,236,326 & 5,132 \\
Nuance & 12 & Publishing &  \$ 2,071,079 & 2,435 \\
TRAX & 10  & Publishing &  \$ 6,509,400 & 4,433 \\
Yral & 16  & Social Media &  \$ 7,540,204 & 16,452 \\
Seers & 13 & Social Media &  \$ 5,371,027 & 2,315 \\
Catalyze & 14 & Social Media &  \$ 5,027,497 & 4,250 \\
BOOM & 14 & Gaming &  \$ 2,490,307 & 5,411 \\
ICGhost & 15 & Meme Coin &  \$ 17 & 2,614 \\
ICPanda & 6 & Meme Coin &  \$ 40,525 & 1,531 \\
Kinic & 16 & AI &  \$ 9,025,317 & 4,918 \\
ELNA & 7 & AI &  \$ 6,134,524 & \hspace{0.4cm} 5,089 \\
\hline
\end{tabular}
\label{table:DAO_overview}
\end{center}\vspace{-.4cm}
\end{table}

The treasury and neuron data was collected from the IC dashboard.\footnote{\url{https://dashboard.internetcomputer.org/sns} on September 30, 2024, using exchange rates from SONIC, a DeFi DAO in the ICP ecosystem }
SNS DAOs vary widely in treasury size, depending on their use case, ranging from \$ 17 for the ICGhost Meme Coin to over \$41 million for ICLighthouse, a DeFi DAO. 
The number of neurons in the ecosystem are also highly variable, with OpenChat and Yral having large neuron populations (around 29,497 and 16,452 neurons, respectively), whereas smaller DAOs like ICPanda have only about 1,531 neurons.

We extracted proposal data for each SNS DAO from the submission of the first proposal until 30.09.2024 by directly querying the governance canisters. 
We evaluated governance dynamics across SNS DAOs by collecting proposal data for each submitted proposal within each SNS DAO. The data for a proposal include the voting outcomes, significant dates and decisions as well as the type of a proposal. For every proposal we defined and calculated key metrics per SNS: approval rate as the proportion of voting power that approved a proposal to total participating voting power, decision-making duration as the time between proposal creation and adoption, and participation rate as the fraction of voting power engaged relative to the total voting power registered. 

\subsection{Participation}
\begin{figure}
    \centering
    \includegraphics[width=\columnwidth]{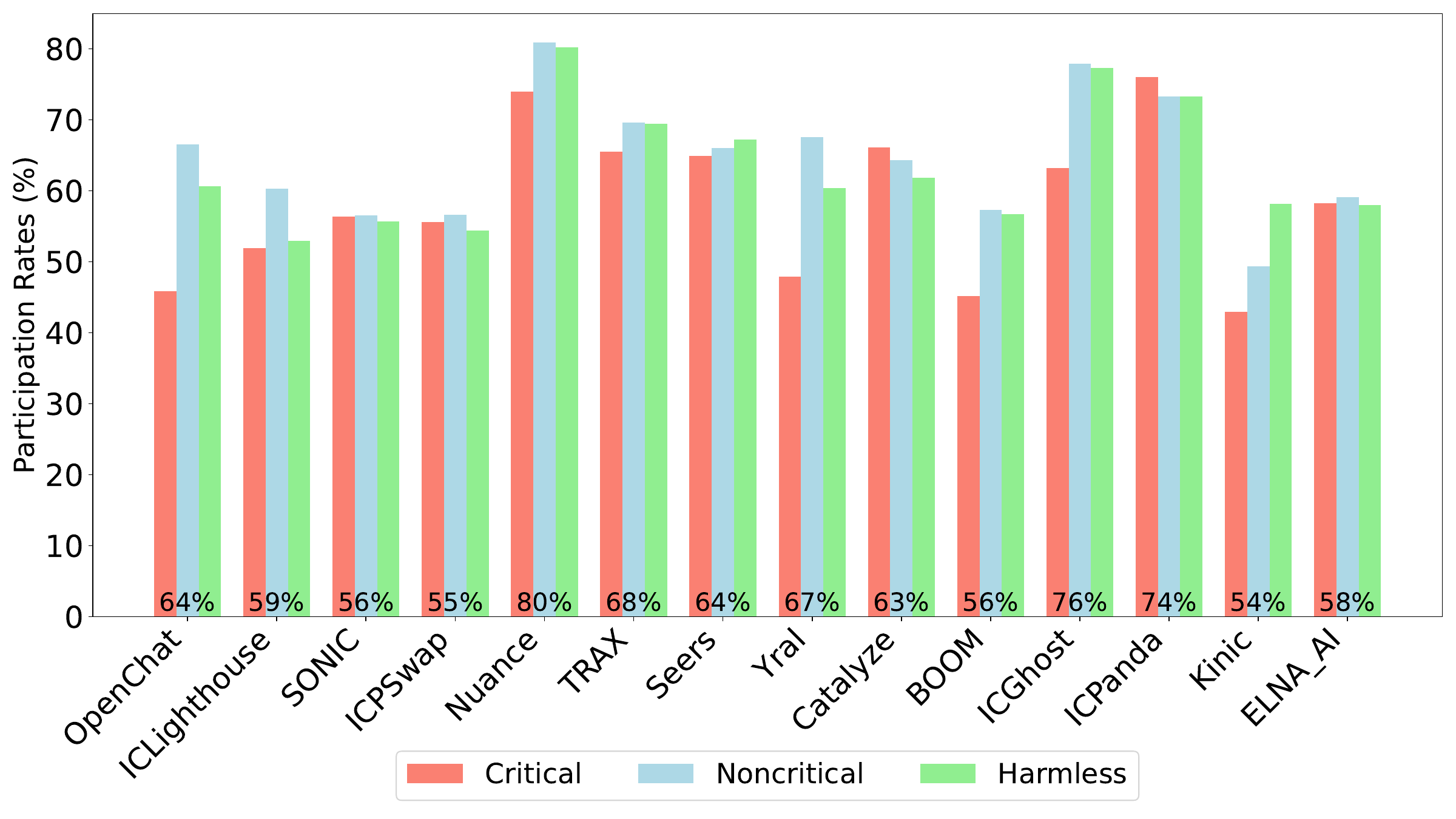}
    \caption{Participation Rates: average proportion of the voting power participating in voting. High rates indicate high community engagement.}
    \label{fig:participation}
    \vspace{-.4cm}
\end{figure}

The governance participation rate measures the ratio of the voting power of the neurons participating in a proposal to the total voting power of an SNS. This forms an essential metric for understanding engagement levels. Low participation rates indicate a passive community
The average participation rate for the SNS DAOs is approximately 64.34\%, see Figure~\ref{fig:participation}. Nuance has the highest participation rate of approximately 80\%. This high level of engagement may be because of the DAO's smaller number of total neurons (2,435). This can encourage a more active community where members feel a stronger responsibility toward governance decisions. However, decisions can also be influenced easier by a few influential neurons since there are less participants to balance the governance decisions. Other SNS DAOs with a significantly smaller number of neurons such as ICPanda and ICGhost are the only ones that are relatively close to Nuance, e.g. 74\% and 77\%. In contrast, Kinic with around 5000 neurons shows the lowest engagement with a rate of around 54\%.

The oldest SNS DAO investigated, OpenChat and Yral, shows moderate rate of participation of voting power, e.g.  64\% and 67\%. Since this SNS DAO is older, it also has more proposals, especially OpenChat with 966 collected proposals, and a higher number of neurons. 
On the other hand, SNS DAOs with 6-7 months such as ICLighthouse, ICPSwap, and ELNA, 6 months, show lower participation rates between 56\% and 59\%. Participation rates do not depend on age only, as demonstrated by ICPanda (also 6 months) and a rate of 74\%. 

 The fact that harmless proposals attract the lowest voting power rate in general indicates, that the community is interested in decisions that impact the future of the project more. However, the participation rates for critical proposals typically lower than for other types of proposals. This probably mostly due to the fact that critical proposals require more effort from the neurons to set up following than for the other categories.

DAOs with higher participation rates may experience more efficient governance processes and better represent the interests of their stakeholders. However, increased participation can sometimes slow down decision-making if not managed efficiently, particularly in DAOs with a high volume of proposals. 
As we will see later there is no definite connection between the participation rates and decision-making durations for SNS DAOs according to our evaluation. 


\subsection{Voter Approval}
Voter approval rates are calculated as the proportion of voting power that supports a proposal relative to the total voting power that is participating in the proposal. This metric provides insight into the collective trust and alignment on similar directions. Higher approval rates can indicate a general consensus for certain governance directions and lower rates may indicate divided opinions or disagreements about proposed changes. 

\begin{figure}[t]
    \centering
    \includegraphics[width=\columnwidth]{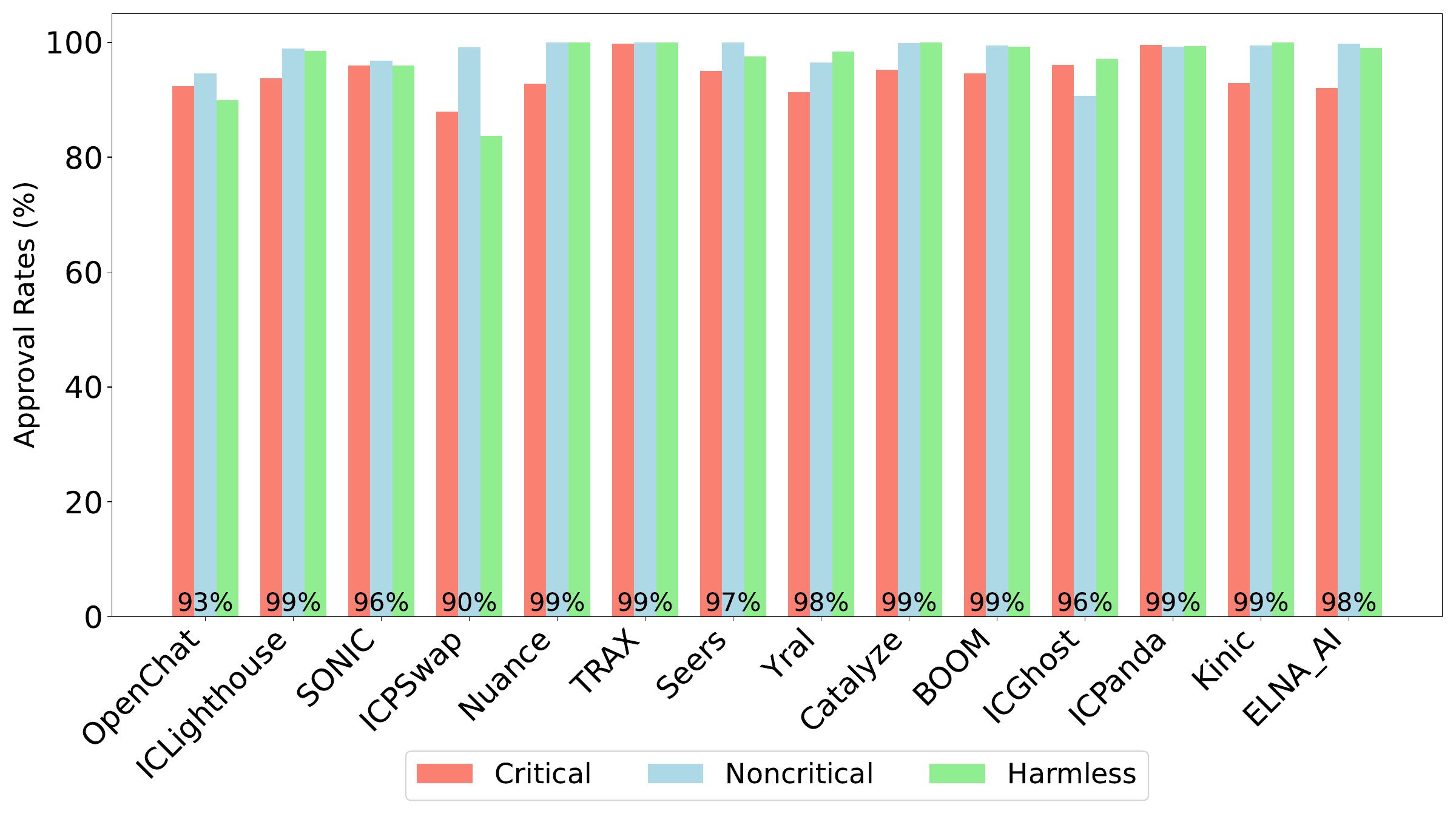}  
    \caption{Approval Rates: average percentage of voting power in favor of adoption of proposals indicating strong consensus among the participants.}
    \label{fig:approval}\vspace{-.4cm}
\end{figure}

The average approval rate across SNS DAOs is approximately 96.8\%, showing a high degree of alignment, with individual approval rates of each SNS DAO visualized in Figure \ref{fig:approval}. Nuance and TRAX, publishing DAOs, exhibit exceptional approval rates of approximately 99\%. Such high rates may imply a close alignment between the proposers and the community or a lower level of diversity in the opinions of stakeholders. This is beneficial for governance efficiency but might also suggest limited evaluation and discussion by participants. Many SNS DAOs with neuron counts below 5000, exhibit rather high approval rates. These high rate of approval could imply that their communities are tightly aligned , possibly due to its smaller, more engaged user base.

ICLighthouse, ICPSwap, and ELNA belong to the set of younger SNS DAOs (6-7 months). They exhibit approval rates of 98.5\%, 90.9\%, and 98.2\%. Interestingly, another one of the younger SNS DAOs, ICPanda (6 months), reveals one of the highest approval rates at approximately 99.3\%. This high level of approval within a small number of neurons suggests strong alignment or dominance by a few neurons with high voting power, which requires further research.

Approval rates are also influenced by the proposal types, e.g., Motion proposals are rejected and disagreed with more often. Therefore, average approval rates are dependent on the proposal dataset and the proportion of the proposal types. 
In some SNS DAOs, such as ICLighthouse, OpenChat, ICPSwap and Seers, approval rates of non-critical proposals are slightly higher than other categories, suggesting a high degree of consensus on mostly operational and less impactful decisions by non-critical proposals. SNS DAOs like Kinic, Nuance and BOOM reveals more varied opinions on critical proposals that could impact the future of the project. This lower approval rates of critical proposals could indicate better deliberation processes and may indicate the awareness of the community regarding impactful decisions, as they mostly need more discussions and time for proposal adoption. For some SNS DAOs, approval rates of harmless proposals are also observed slightly lower than other categories, such as ICPSwap and OpenChat. This is probably due to the contextual and easily rejectable nature of “Motion” proposals that often consist of discussions and suggestions.

\subsection{Rejection Trends}
\begin{figure}
    \centering
    \includegraphics[width=\columnwidth]
{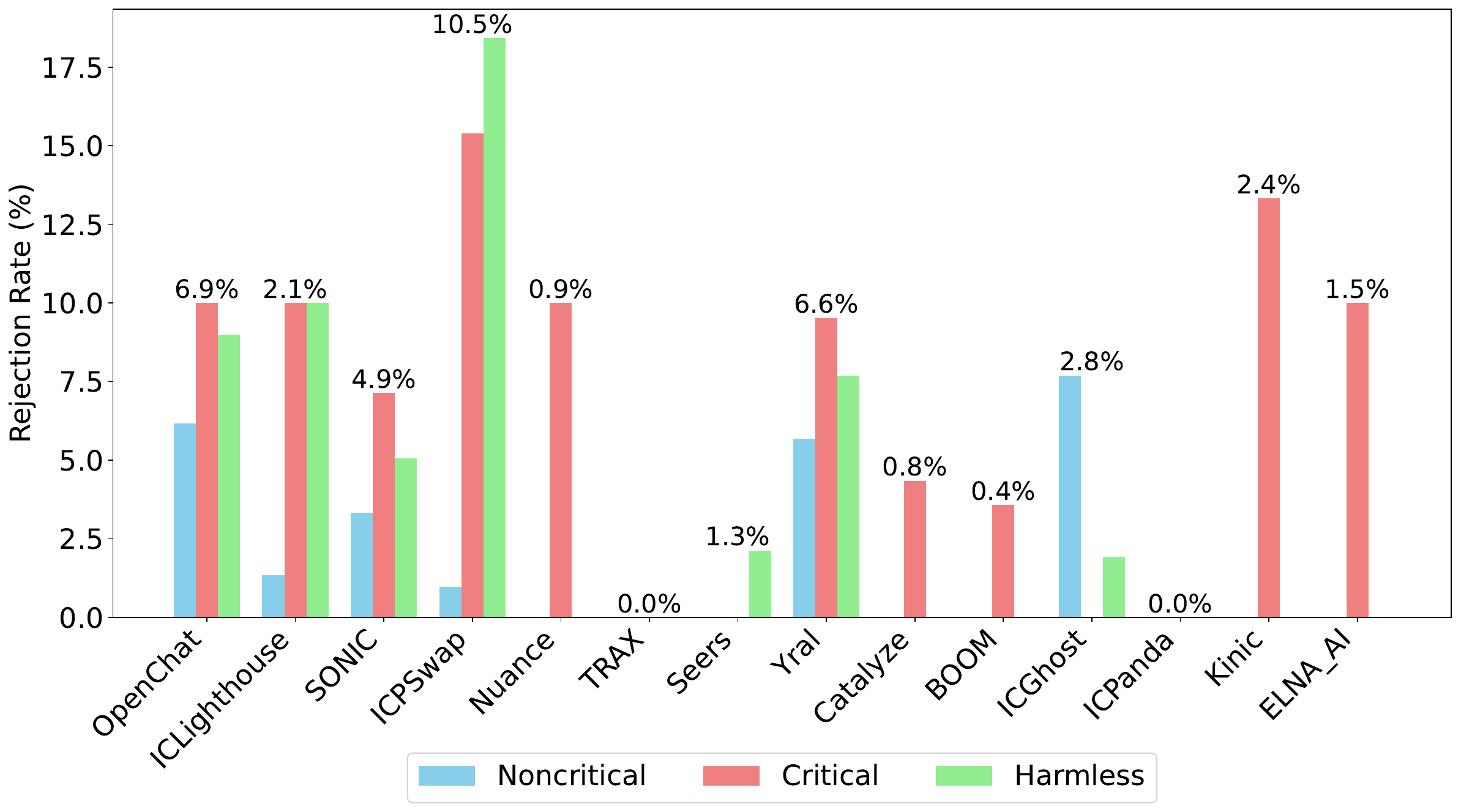}
    \caption{Percentage of rejected proposals, depending on their category.}
    \label{fig:rejection}\vspace{-.4cm}
\end{figure}

While most proposals are adopted with a high voting power percentage, some proposals do get rejected. A low fraction of non-critical proposals are rejected (less than 8\% in all SNSs, above 5\% in 3 SNSs only), while in some SNSs more than 15\% reject harmless proposal as we can see in Figure \ref{fig:rejection}. With the exception of ICSwap (15\%) and Kinic (13\%), all SNSs reject fewer than 10\% of the critical proposals, yet more than non-critical proposals. This indicates that community members approach these high importance proposals more cautiously. Non-critical proposals are the ones that are least rejected, probably due to their lower impact and the large number of non-critical proposal types, that are proposed more often. The relatively high number of rejected harmless proposals can explained by the nature of “Motion” proposals which query the SNS community for their opinion without an automatic execution of any code upon adoption.
SNS DAOs such as Kinic, Nuance and ELNA, exhibit interesting results regarding critical and harmless proposals. In their proposal dataset, there is not a single rejected non-critical or harmless proposal, but instead reject more critical proposals than other SNS DAOs. This could be due to community alignments or total number of proposals, which are relatively small for each of them. 

In order to evaluate the rejection trends better, we have also investigated which specific types of proposals are mostly rejected, regarding the critical and harmless proposals. We have identified two interesting patterns depending proposal types, e.g. “Motion” and “TransferSNSTreasuryFunds” proposals. Harmless proposals consists of “Motion” and “SNS-Upgrade” proposals. Out of all 14 SNS DAOs we have not identified a single case where a “SNS-Upgrade” proposal is rejected. High rejection rates of harmless proposals therefore only stem from “Motion” proposals. As we have already elaborated, “Motion” proposals are a means for community discussions and placing suggestions suggestions, or even a former discussion of a feature extension. This means that they are more likely to be rejected and disagreed upon, because these proposals are specifically submitted to hold discussions and gather diverse opinions. Secondly, we have identified that rejection rates of critical proposals all depend on a single proposal type again, which is the “TransferSNSTreasuryFunds” proposal. We did not identify a single case again, where another type of critical proposal is rejected. This proposal type requires complex calculations and has a great impact on the ecosystem, as the SNS DAOs control their treasury and transferring of funds to other accounts could be disagreeable and controversial, therefore leading to higher rejection rates.

\subsection{Decision-Making Durations}

The decision-making duration is the time taken between the creation and adoption of a proposal. 
It is an important factor that reflects the responsiveness and governance efficiency of SNS DAOs. Shorter durations are useful in adapting quickly to changes. Longer durations on the other hand may reflect inefficiencies in processing proposals but also indicate that neurons are more inclusive in the decision-making and the results could be more reflective of the community. 

\begin{figure}
    \centering
    \includegraphics[width=\columnwidth]{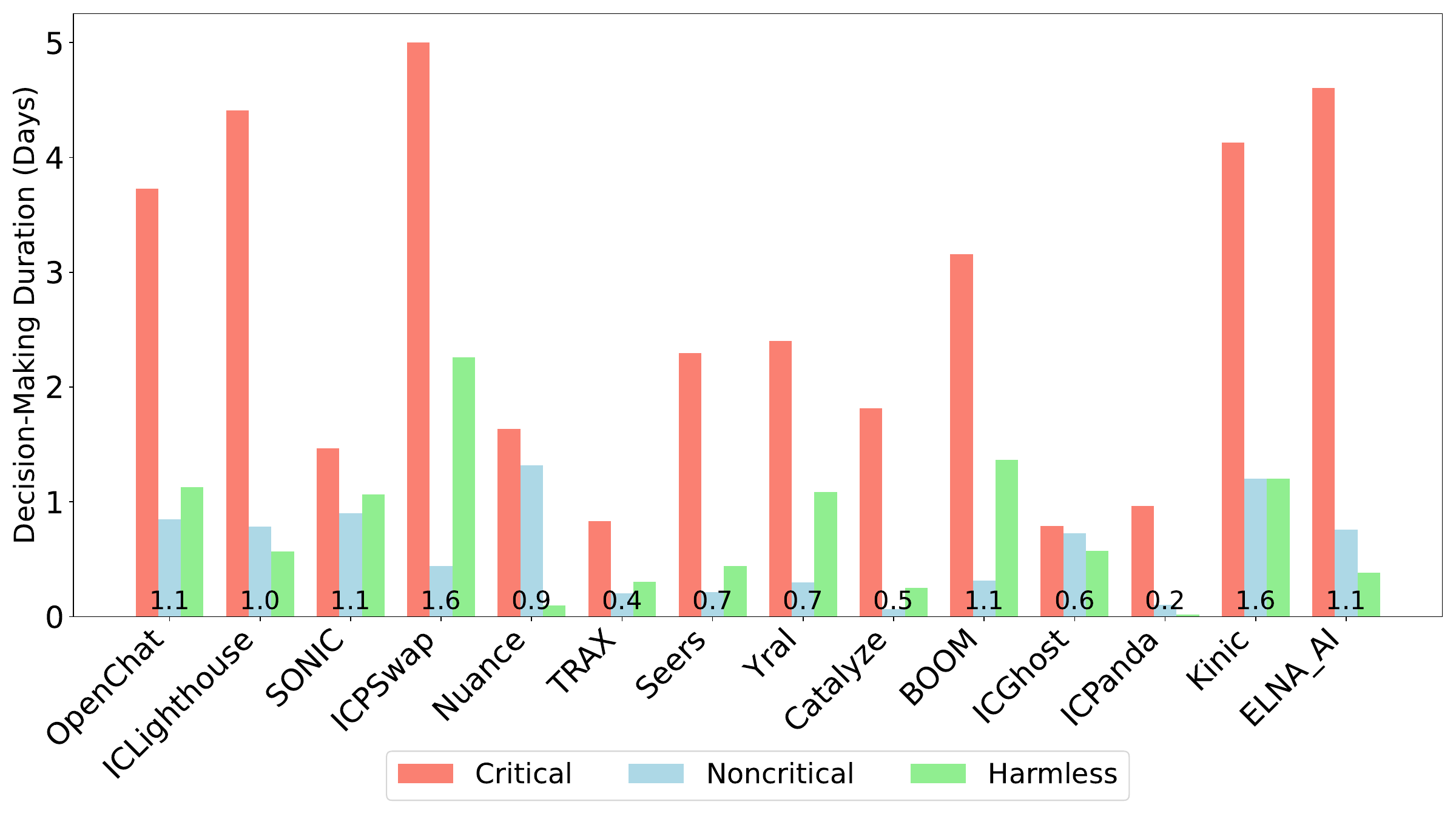}
    
    \caption{Decision-Making Durations: Average time between proposal submission and decision. Short durations indicate community responsiveness and depend on the proposal types, critical proposals imply longer durations.} \vspace{-.4cm}
    \label{fig:durations}
\end{figure}

Decision-making durations vary significantly between SNS DAOs and between proposal types, see Figure \ref{fig:durations}, with an overall average duration of approximately 1.14 days.

Critical proposals require significantly longer time in all SNS DAOs, most likely because the voting period is longer for critical proposals, as they require more attention and engagement. For example, in ICLighthouse critical proposals are decided in approximately 4.4 days, compared to non-critical proposals with 19 h. In some of the SNSs, harmless proposals also exhibited a higher duration that non-critical proposals, probably due to “Motion” proposals that are more frequently rejected and encourage discussions due to their nature.

ICPanda and ICGhost have short average decision-making duration, averaging around 3.84 h and 14.8 h,  possibly due to their category as a meme coin. Quick, community-driven decisions could be favored, because there is no need for complex improvements and constant feature developments. 
ICPanda and ICGhost show a relatively low ratio of critical proposals to total proposals (e.g. 9\% and 7\%) , which indicates that significantly lower average decision-making durations may be influenced by a small fraction of critical proposals in the dataset. Similarly, TRAX has relatively low fraction of critical proposals, with 16\% of total proposals, and it also is one of the fastest SNS DAOs with an average of approximately 8.5 h. However, it is important to mention that average durations for critical proposals for these 3 SNS DAOs are also the lowest ones, influencing the overall lower durations.

SNS DAOs such as Kinic and ELNA, both related to artificial intelligence, with a slightly higher fraction of critical proposals (e.g. 13\% and 14\%), exhibit much higher average decision-making durations, e.g. 1.59 (highest average duration) and 1.1 days for a proposal. Therefore, although influential, ratio of critical proposals to total proposals is not the only factor that influences average decision-making durations in SNS DAOs, e.g., complex decision-making processes and proposals that needs expertise and information handling. 

DeFi DAOs, ICLighthouse (23.94 h), SONIC (25.73 h) and ICPSwap (1.58 days), exhibit similar average decision-making durations.  Since their total number of neurons and number of critical proposals 
are similar, longer durations could indicate the complexity of decision-making processes and features due to their category and partially the impact of having added many application-specific proposals. 
ICLighthouse features 97 additional proposal types, ICPSwap has 10, more than most of the other SNS DAOs, while SONIC has added 2 additional proposal types. 
Neurons may lack information on these types and therefore extend the duration. OpenChat, the oldest SNS, also has a similar average decision-making duration with 25.5 h and also exhibits a large number of 41 additional proposal types (the median over all DAOs is 3). 

Our evaluation on decision-making duration revealed variety of average durations across all SNS DAOs. For SNS DAOs with frequent proposals, efficient decision-making processes are essential for ensuring responsive governance structures that prioritizes development speed. 
Categorization of critical and noncritical proposals and decision-making rules based on them by SNS DAOs could serve as an example for maintaining this balance and could be further evaluated for different governance models and optimization.

\subsection{Time-Based Evaluation of Governance Dynamics}
In addition to the static metrics considered so far, we are also interested in how governance dynamics evolve over time, especially with respect to community engagement, as other studies revealed a decline in voting activities. 

\subsubsection{Participation and Approval}
Our evaluation on the changes of participation and approval rates over time shows small variations across different SNS DAOs. In most of the SNS DAOs both rates show small fluctuations and participation rates show an overall increase, while approval rates are more consistent.

\begin{figure}
     \centering
         \includegraphics[width=\columnwidth]
         {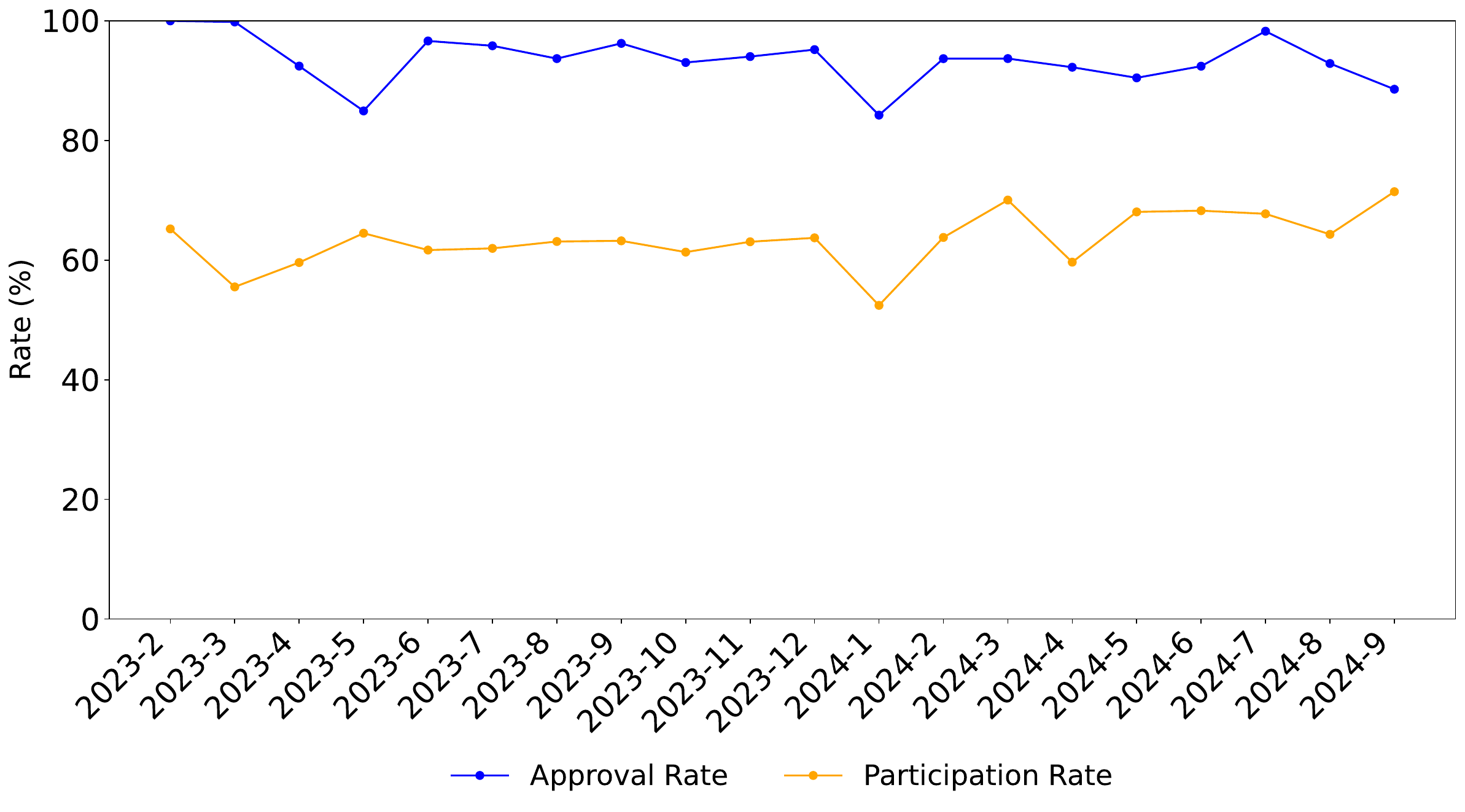}    
         \caption{Monthly average participation and approval rates for OpenChat, revealing small fluctuations but consistent rates}
         \label{fig:shifts_chat}\vspace{-.4cm}
 \end{figure}%

OpenChat, for example, exhibits a relatively low participation rate of 52.4\% in Jan 2024, followed by an increases, with small fluctuations up to 71.4\% in Sep 2024. Overall, both participation and approval rates fluctuate $\pm$ 10\%, which may be due to variance in neurons and their behavior as well as the impact of proposals in specific months.

\begin{figure}
    \centering
        \includegraphics[width=\columnwidth]
        {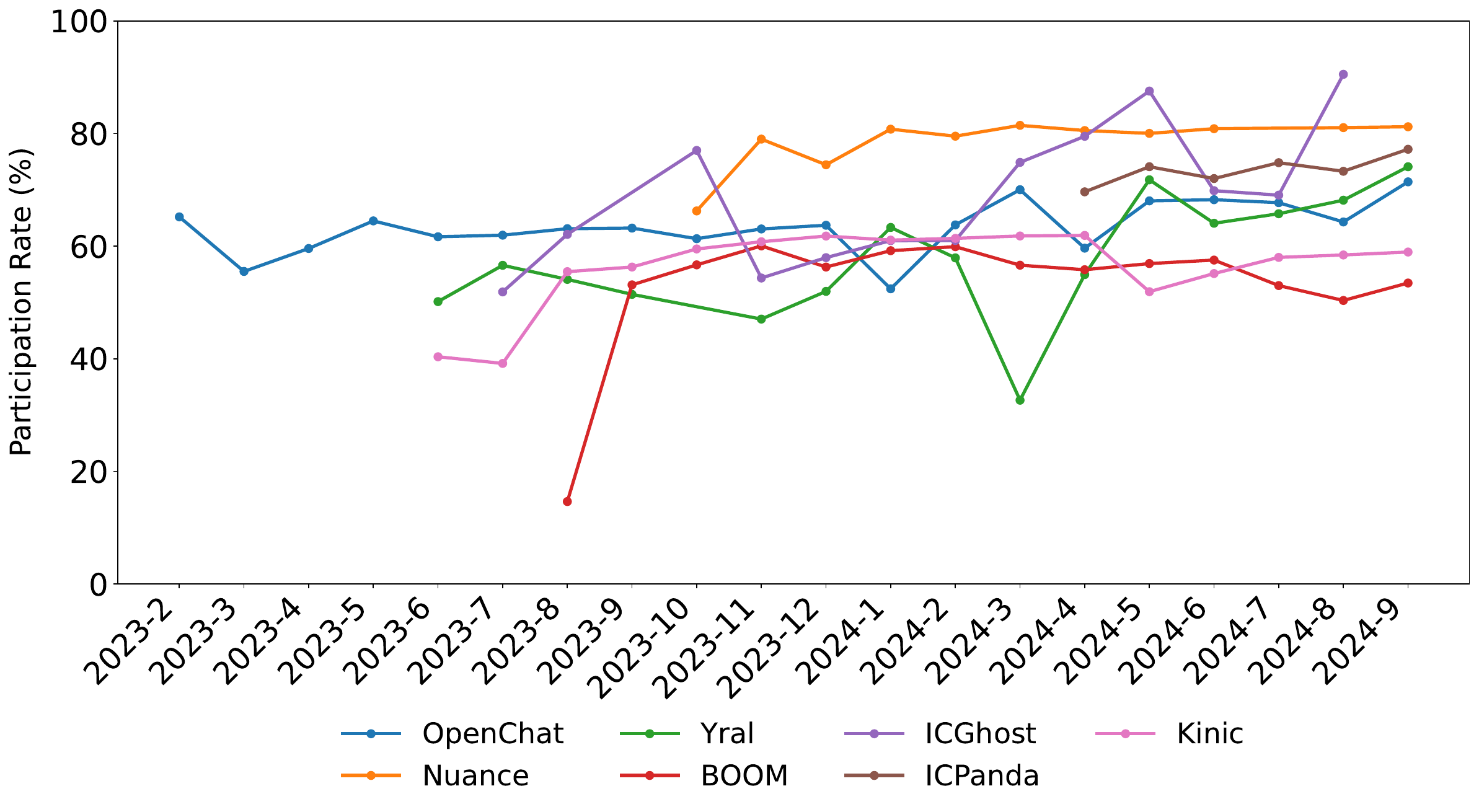}    
        \caption{Monthly average participation rates for some SNS DAOs, showing fluctuations with an overall upwards trend}
        \label{fig:shifts_combined}\vspace{-.4cm}
\end{figure}%

All SNS DAOs reveal similar results to OpenChat, see ~\ref{fig:shifts_combined} for a plot of a subset of participation trends. Participation rates encounter some fluctuations but as of September 2024, all of them are higher than at launch. Approval rates are generally very high and do not change significantly. 
In general, younger SNS DAOs such as ICLighthouse, ICPSwap and ICPanda exhibit more consistent rates. All of these SNS DAOs are founded after March 2024, where it could be expected that governance structure and dynamics of SNS DAOs have been understood and optimized more compared to older SNS DAOs. 




\subsubsection{Monthly Proposal Frequencies}


Examining the frequency of proposals as well as the monthly shifts across SNS DAOs provides insight into the activity levels within each community and their governance processes. 

\begin{figure}[b]\vspace{-.4cm}
    \centering
    \includegraphics[width=\columnwidth]{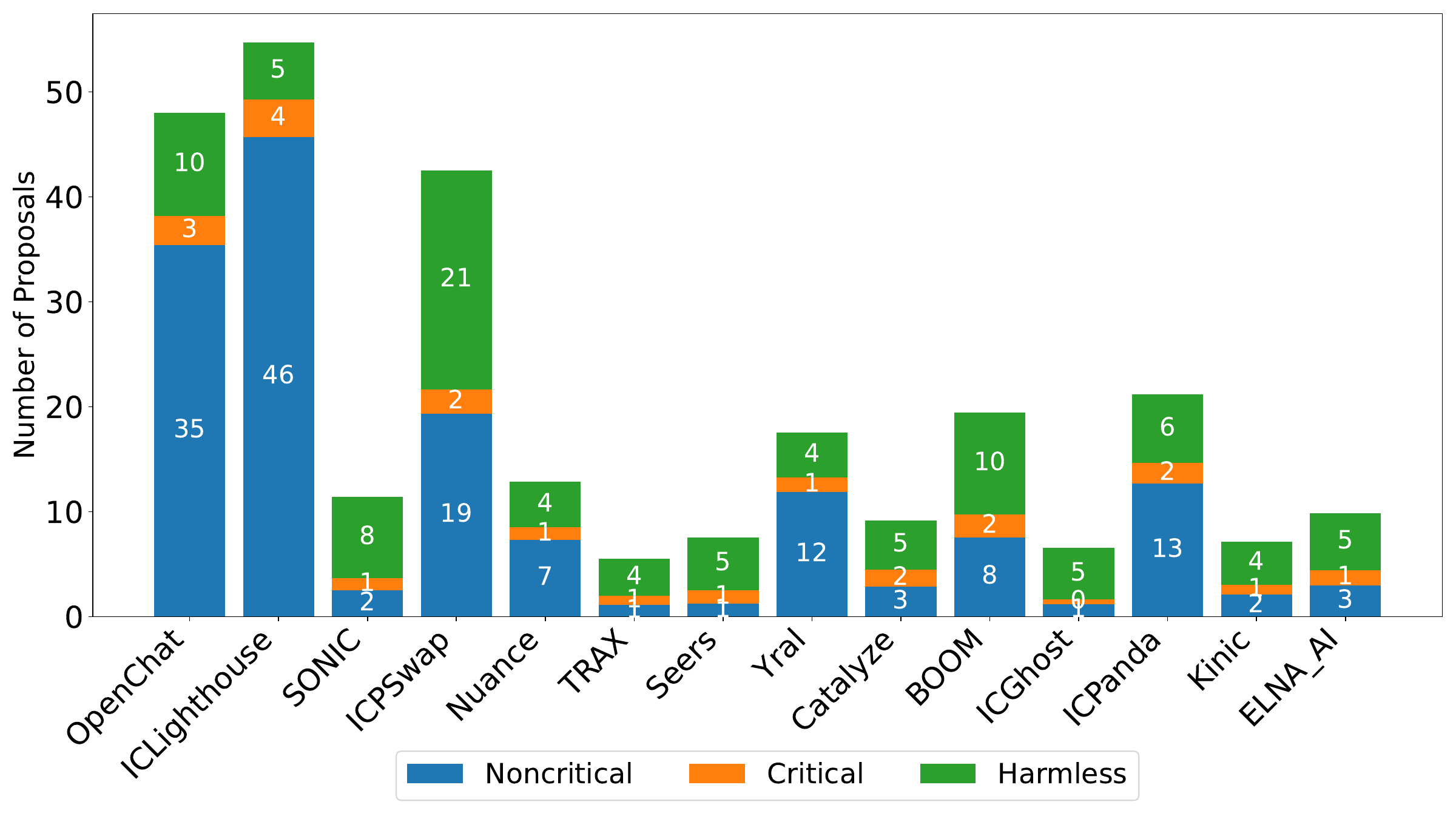}    
    \caption{Average monthly proposal frequencies, rounded to integers. Some SNS DAOs are much more active than others.}
    \label{fig:frequency}
\end{figure}

Figure \ref{fig:frequency} shows the average number of proposals that are submitted in a month for each SNS DAO. OpenChat, ICLighthouse adn ICPSwap have the highest average proposal frequencies with 48.3,  54.86  and 42.5 monthly proposals. Other SNS DAOs do not exhibit a similar proposal activity with ICPanda having the highest rate among them, 21.17 proposals per month. Factors such as age, number of neurons and the voting power distribution do not play a major factor in monthly proposal activity. ICPanda, with the highest number of monthly proposals among SNS DAOs, except for OpenChat, ICLighthouse and ICPSwap, has the lowest number of neurons across all SNS DAOs. Proposal activity could likely be influenced by the communities' interests, need for development and management. Higher proposal frequencies, as seen in OpenChat, could indicate active governance and an environment for continuous development. DAOs with lower frequencies, like TRAX having the least number of average proposals with 5.5 per month, could potentially lead to members seeking more active participation losing their interest over time.

\begin{figure}
    \centering
    \includegraphics[width=\columnwidth]{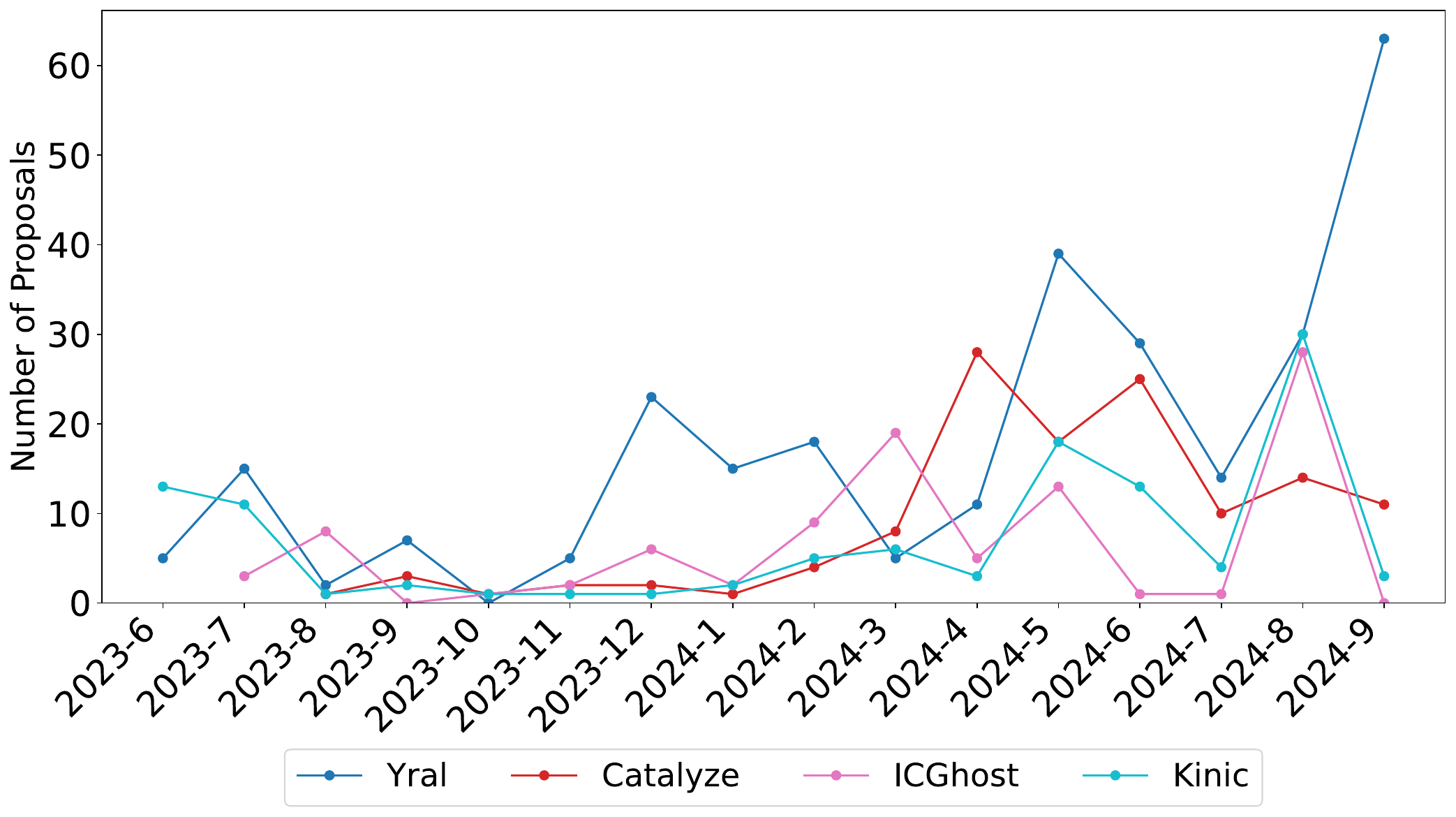}
    \caption{Average monthly proposal frequencies of a subset of DAOs.}\vspace{-.4cm}
    \label{fig:combined}
\end{figure}

Another factor that could influence the level of proposal activity is the cost for government. Submitting a proposal requires substantial effort that could potentially hinder the engagement of neurons, that do not have a great token value and voting power. Therefore, as we have seen in the evaluation of proposer varieties, only a small number neurons actually propose a proposal and share their opinion. Having a larger set of proposers could lead to diverse opinions and therefore more frequent proposals. However, balancing the cost and governance efficiency is a crucial factor here, as an extensive number of proposals could hinder the governance engagement, if most of them lack expertise or are spams. In order to fully understand the reasons behind the variety of proposal frequencies, further research on governance costs, as well as community interests and trends should be conducted. 


Our evaluation on the monthly changes of number of proposals revealed a variety of fluctuations for most of the SNS DAOs without a consistent increase or a decrease in numbers, as exemplified by four SNSs in Figure~\ref{fig:combined}. Other SNSs show similar numbers, while most of them showing an increase in submitted proposals, especially in the last months, compared to initial months of creation. The number of proposals proposed in a month can be influenced by various factors, such as the voting behavior of influential neurons in that particular time, community needs and developments during certain events as well as significant changes in governance structures that needs further research.


\section{Comparative Analysis of Governance Metrics Across DAO Platforms}
\label{sec:comparison}

In this section, we analyze and compare the governance metrics observed for SNS DAOs on ICP with those from DAOs operating on other blockchain platforms \cite{feichtinger2023,messias2023,barbereau2022.2,wang2023,faqir2021}. The aim is to evaluate the governance dynamics of SNS DAOs within the broader landscape of decentralized governance and highlight their unique strengths and challenges. 

\textbf{Participation rates}, a key indicator of voter engagement and inclusiveness, are relatively high in SNS DAOs, averaging at approximately 64\% with variations exist across DAOs and proposal categories.
In comparison, participation rates of voting power from Ethereum-based DAOs, such as Compound (34\%), UniSwap (31.4\%), ENS (39.2\%), Gitcoin (28.6\%), are much lower than SNS DAOs \cite{feichtinger2023}. Messias et al. also investigated the UniSwap and Compound for their voter engagements by calculating the ratio of the number of votes (or delegated tokens) cast on a proposal to the total number of delegated tokens \cite{messias2023}. Their results show that the average participation rate is 33.25\% and they observe that earlier proposals have higher engagement rates. In addition Rhazoui et al.~investigated DAO platforms such as Aragon, DAOstack and DAOhaus in terms of voter engagement, revealing that voter participation is below 10\% on average for DAOstack and Aragon, while DAOhaus reached 40\% \cite{faqir2021}. Furthermore, to give a comparison, Feichtinger et al. calculated the participation rates for UniSwap and Compound, which are much lower than the participation in voting at shareholder meetings in traditional companies of US (70\%) \cite{feichtinger2023}. 
Barbereau et al. investigated voting rights and power distribution as well as voting activity of 9 different Ethereum-based DeFi DAOs, including Uniswap and Compound \cite{barbereau2022.2}. 
They examined the exercised voting rights over time, typically less than 1\% of their token-holders engaged. They also reveal that participation rates observed in earlier months do not last long and even for the DAOs with greater engagement levels such as Yearn Finance these rates decreased drastically over time. In contrast, SNS DAOs do not change much over time and for most cases the participation rates are higher than the initial months.

Hence, SNS DAOs have a greater voter engagement than other DAO ecosystems. Compared to other platforms, SNS DAOs benefit from the fact that tokenholders do not have to pay for submitting their votes or selecting followees and they are rewarded for these activities by most SNS DAOs.


\textbf{Approval rates} of SNS DAOs reflect a high degree of consensus, with some "Critical" proposals, that have bigger impact on the future of the DAO, achieving a lower approval rates and indicating a small diversity of opinions within the community. The average approval rate across SNS DAOs is approximately 96.8\%, showing a high degree of alignment, with most SNS DAOs showing rates above 90\%.  

Messias et al. reported similar results to ours while evaluating the Compound and Uniswap protocols. The majority of the proposals in Compound received significant support with an average of 89.39\% in favor votes \cite{messias2023}. In their research, Rhazoui et al. revealed similar results, with 86\% of positive votes among those cast for DAOstack, 91\% for DAOhaus and 94\% for Aragon \cite{faqir2021}. Wang et al. explore a large dataset of DAO projects and off-chain votings collected from Snapshot \cite{wang2023}. They investigated how much a proposal is being agreed with or opposed by the community and found that over 60\% of proposals feature a difference of more than 40\% between the positive and negative votes. 

These findings position SNS DAOs alongside Ethereum DAOs in fostering consensus-driven decision-making, while showcasing a slightly broader diversity of opinions for critical categories. Our results suggest that SNS DAOs have a greater community consensus than most of the DAOs in other blockchain platforms, probably due to fact that they are smaller in terms of eligible voters and most proposals are of high quality without hardly any spam. 

\textbf{Decision-Making Durations} for SNS DAOs are relatively short, reflecting streamlined voting mechanisms. This efficiency is particularly notable when contrasted with other blockchain platforms. The overall average decision-making duration for SNS DAOs is approximately 1.14 days (27.4 h). 

While investigating 581 DAO Projects and off-chain votings results from Snapshot, Wang et al. \cite{wang2023} observed longer intervals to reach decisions. They report that 13.8\% of the proposals are decided in more than a week, 25.7\% of proposals are decided in 3 days, 14.2\% of proposals in 2 days and  only 3.2\% are decided in less than a day. Messias et al. calculated durations for Compound and UniSwap and observe durations of 1.64 days to reach the quorum in average \cite{messias2023}.  

Our results on the durations of proposals for SNS DAOs are much lower than \cite{wang2023} but similar to Compound and Uniswap \cite{messias2023}. Many SNS DAOs take less than a day to reach a quorum on average. Voting periods and thresholds for “critical” proposals are intentionally higher, therefore increase the results for average durations, to give the community enough time to evaluate their decisions on the impactful proposals, which take much more time than other proposal types. Non-critical proposals are mostly decided within a day reflecting the efficiency of governance mechanisms and processes within SNS DAOs. With the flexibility offered by adjusting voting periods depending on the importance of proposals, SNS DAOs provide an example framework for efficient decision-making mechanisms while also allowing tokenholders to discuss and take their time on important decisions.

\textbf{Governance activity} is reflected by the frequency of proposal submissions.  OpenChat (48)  and ICLighthouse (55) have the highest average monthly proposal frequencies.  Other SNS DAOs do not exhibit a similar proposal activity but are mostly between 5-10 per month. TRAX has the lowest number of average proposals with 5.5 per month.   

Messias et al. \cite{messias2023} count a maximum number of proposals created in a month of 11 and 10 when investigating the activity levels of Compound and Uniswap, with an average of 4.3 and 1.7 proposals per month on average. The longest interval between proposals was 31 and 81 days. When studying Uniswap more closely, the maximum activity occurred in the first month, never reached again later.

In comparison, SNS DAOs show considerably higher numbers of proposals per month, indicating a more active community and interest in development. For most of the SNS DAOs, especially older and more active ones, our findings reveal that proposals are submitted almost every month for most SNS DAOs without a huge interval without activity. For SNS DAOs such as OpenChat and ICLightouse proposals are submitted almost every day and more than one proposal in some cases. When examining the monthly changes in proposal frequencies for SNS DAOs, no patterns emerge, thus we conclude that frequencies fluctuate depending on the needs and interests of the community and SNS DAO development. 


\textbf{Governance cost and rewards}
Another factor is the cost for governance in terms of time and money. Submitting and voting actively on a proposal requires effort and depending on the DAO framework the monetary cost is not negligible. Creating a proposal is more involved than delegation or voting, which explains the small number of proposers observed in most DAOs.  Different sources report average costs for a compound proposal are 594.17 USD in total (7.88 USD per vote)~\cite{messias2023} as well as numbers between 2500 USD and 20,000 USD per proposal for Gitcoin, Uniswap and Compound, depending on the overhead costs considered~\cite{feichtinger2023}.
In contrast, the cost for all SNS canisters (including functionality beyond voting, e.g., ledger transfers or upgrading canisters) divided by the number of executed proposals in December 2024 is around 11 USD, with an average transaction cost of 0.000027 USD (on average 180 replicated operations per second are executed for SNS DAOs).
SNS DAOs’ activity levels, especially the monthly proposals frequencies, participating and approving voting power are higher than the DAOs analysed from other frameworks. The fact that for SNSs voting does not require spending tokens and setting up followees is fairly easy is a likely explanation for the sustained or even increasing activity across a diverse range of proposals observed for SNSs. 

SNSs promote participation by distributing rewards in SNS-native tokens, proportional to a neuron's voting power. Voting power increases with locking period and neuron age, encouraging long-term commitment and minimizing short-term manipulation. Numerous studies confirm that incentive design plays a critical role in DAO governance performance. \cite{Piyarisi2025} demonstrate that effort and engagement in DAOs scale directly with token-based compensation, particularly when rewards are proportionate to governance involvement. Their model shows that decentralized organizations employing effort-dependent reward structures—such as those in SNS DAOs—achieve greater participation. SNS’s implementation of voting power and rewards aligns with this principle. \cite{Davidson2023Compensation} further emphasizes the inadequacy of flat token-weighted voting models, suggesting role-sensitive, recurring compensation systems that reflect actual engagement. SNS DAOs achieve this by dynamically rewarding governance activity and incentivizing long locking periods. \cite{murano2024incentive} highlight that voting costs and low expected utility often disincentivize participation—a dynamic countered in SNS DAOs by the reverse gas model and following.

These findings validate and correlate with high proposal frequency, engagement and strong voter alignment. Unlike Ethereum-based DAOs, where gas costs and the lack of token based dynamic incentive mechanisms can discourage participation, SNS fosters a structurally inclusive, economically rational environment for governance activity.


\textbf{Summary and Evaluation} This comparison highlights the strengths and challenges of SNS DAOs within the broader DAO ecosystem. They excel in participation, proposal activity, and agile decision-making. Opportunities for improvements remain to enhance voter incentives and engagement levels especially for critical proposals that have a bigger impact on the future of the DAO. By building on these insights, SNS DAOs can serve as a model for scalable and efficient decentralized governance. SNS DAOs provide valuable lessons for other blockchain ecosystems. Their on-chain frontend simplifies participation. The low proposal cost and reverse gas model reduce participation barriers seen in Ethereum DAOs where fees can exceed thousands. Topic-specific, majority-based delegation implements a robust form of liquid democracy, empowering users to delegate votes by expertise. Voting power increases with locking duration and age, incentivizing long-term commitment. Voters are rewarded in SNS-native tokens, aligning engagement with ecosystem growth. This structure creates consistent participation, high alignment, and a more resilient governance process. Despite their benefits, adopting SNS-like mechanisms elsewhere faces several challenges such as high gas fees on Ethereum, dependency on off-chain platforms, and lacking of flexible, long-term incentive systems that weakens engagement. SNS DAOs offer a replicable design, but require compatible infrastructure and incentive-aligned protocols to achieve similar outcomes.

\section{Conclusion}\label{sec:conclusion}

We presented an empirical study of the efficiency and activity
of SNS DAO governance processes on ICP, comparing our results to other frameworks.
We found moderate to high participation levels, depending on the type of DAO, and generally see an increasing trend over time. We also generally observed high approval rates, and relatively fast decision making. This indicates that the SNS mechanisms overcome some of the challenges observed on other DAO frameworks, in particular the studied SNS DAOs do not suffer from a decline in participation but increased engagement. We hope that our work inspires follow-up research, on additional aspects on SNS or other DAO frameworks, e.g., considering more long-term time spans and delegation patterns as well as the power asymmetries among neurons.
\bibliographystyle{plain}
\bibliography{SNS}

\end{document}